\documentclass[aps,prl,twocolumn,groupedaddress,superscriptaddress]{revtex4}
\usepackage[dvips]{graphicx}
\usepackage{empheq}

\newcommand{\ket}[1]{\left|\;#1\;\right\rangle}

\begin{document}
\title{Propagation of the photoelectron wave packet in an attosecond streaking experiment}
\author{E. E. Krasovskii} 
\affiliation{Departamento de F\'{i}sica de Materiales, Facultad de
  Ciencias Qu\'{i}imicas, Universidad del Pais Vasco/Euskal Herriko
  Unibertsitatea, San Sebasti\'{a}n/Donostia, Spain}
\affiliation{Donostia International Physics Center (DIPC),
  San Sebasti\'{a}n/Donostia, Spain}
\affiliation{IKERBASQUE, Basque Foundation for Science, Bilbao, Spain}

\author{C. Friedrich}
\affiliation{Peter Gr\"unberg Institute and Institute for Advanced Simulation, 
Forschungszentrum J\"ulich and JARA, 52425 J\"ulich, Germany}

\author{W. Schattke}
\affiliation{
Institut f\"ur Theoretische Physik und
Astrophysik der Christian-Albrechts-Universit\"at zu Kiel,
Leibnizstra{\ss}e 15, 24118 Kiel}
\affiliation{Donostia International Physics Center (DIPC),
  San Sebasti\'{a}n/Donostia, Spain}

\author{P. M. Echenique}
\affiliation{Departamento de F\'{i}sica de Materiales, Facultad de
  Ciencias Qu\'{i}imicas, Universidad del Pais Vasco/Euskal Herriko
  Unibertsitatea, San Sebasti\'{a}n/Donostia, Spain}
\affiliation{Donostia International Physics Center (DIPC),
  San Sebasti\'{a}n/Donostia, Spain}
\affiliation{%
Centro de F\'{\i}sica de Materiales CFM - Materials
Physics Center MPC, Centro Mixto CSIC-UPV/EHU, Edificio Korta,
Avenida de Tolosa 72, 20018 San Sebasti\'an, Spain
}

\begin{abstract}
Laser-assisted photoemission from a solid is considered within a numerically 
exactly solvable one-dimensional model of a crystal. The effect of the inelastic 
scattering and of the finite duration of the pump pulse on the photoelectron 
dynamics is elucidated. The phenomenological result that the photoexcited wave
packet moves with the group velocity $dE/dk$ and traverses on average the distance  
equal to the mean free path is found to hold for energies far from the spectral 
gaps of the final state band structure. On the contrary, close to a spectral gap
the photoelectron is found to move during the excitation by the pump pulse with 
a velocity higher than the group velocity.
  \end{abstract}

\maketitle
\section{introduction\hfill ~}
Transport properties of electron wave packets underlie the functioning of 
electronic devices and are an important factor in photoemission spectroscopies
and electron diffraction techniques. In a solid, as opposed to a single atom, 
special role is played by inelastic scattering giving rise to a limited electron 
lifetime and mean free path. The inelastic scattering is phenomenologically
described as electron absorption~\cite{Slater37}, and its implications for stationary 
processes are well understood
\cite{Strocov2001,Barrett05,Strocov06,Krasovskii07,Krasovskii08,Krasovskii15}. 
A much less studied question is how the wave packet evolves and propagates on a time 
scale comparable to its lifetime. Experimental access to such ultrafast processes is 
given by modern time-resolved photoelectron spectroscopies, in particular by the 
streaking method~\cite{KI09,Hen2001N,Dre2001S,Kie2004N,Ca07}, in which a subfemtosecond 
time resolution is achieved by mapping time to energy using a strong laser field: 
the electron wave packet created by an ultrashort pulse of extreme ultraviolet 
radiation (XUV) is accelerated by the superimposed laser field, and the energy by 
which its spectrum shifts up or down depends on the electron release time 
$t_{\textsc x}$ relative to the temporal profile of the laser field $E_{\textsc l}(t)$.

Various models have been developed that treated the problem by the time-dependent
Schr\"odinger equation (TDSE)~\cite{BM08,KE09,ZT09,Borisov13} or by applying a 
classical model~\cite{Le09}. The majority of the studies consider a solid in which 
the low-frequency laser field is strongly damped by the dielectric 
response~\cite{Krasovskii10}, so the electron needs to travel some distance
before it gets exposed to the streaking field.

\begin{figure}[b]
\includegraphics[width=0.48\textwidth]{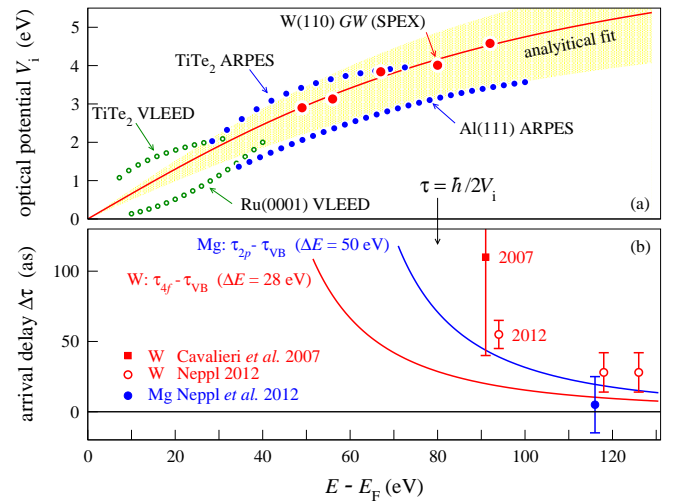}
\caption{
\label{univi} Dependence of optical potential $V_{\rm i}$ on energy.
Small circles are the values empirically derived from {\it ab initio}
analysis of ARPES spectra (full circles) and VLEED spectra (open circles). 
Large circles are the {\it ab initio} $GW$ calculation for W(100) (FLEUR-SPEX). 
The dependence $V_{\rm i}(E)$ is well fitted by a function
$\gamma\arctan(E/\beta)$ with $\beta=80$~eV and $\gamma=5.3$~eV for W(100),
$\gamma=4$~eV for Al(111), and $\gamma=6.2$~eV for TiTe$_2$.
}
\end{figure}
\section{MEAN FREE PATH OR LIFETIME?\hfill ~}
Assuming an instantaneous XUV excitation, the time of flight from 
the depth where the photoelectron is created to the surface is determined by 
the group velocity of the wave packet. Because of the inelastic scattering the 
photoelectrons excited deeper in the crystal have smaller probability to reach 
the detector, which is assumed to scale with the depth $z$ as $\exp(z/\lambda)$. 
Then the average depth traversed by the electrons is just the mean free path (MFP) 
$\lambda$, and the time to reach the surface can be estimated as $\tau=\lambda/v$, 
where $v$ is the group velocity of the emitted wave packet~\cite{Ca07}. Often it 
is sufficient to use the values of $\lambda$ averaged over an energy interval of 
several eVs that can be inferred from the well-known universal $\lambda(E)$ curve. 
This is, however, not sufficient for ARPES because MFP may strongly vary over a 
few eV interval as a result of the specific band structure. At the same time, 
experience with ARPES and VLEED (very low energy electron diffraction) 
\cite{Barrett05,Strocov06,Krasovskii07,Krasovskii08,Krasovskii15}
suggests that the photoelectron final-state lifetime changes rather smoothly with 
energy, see Fig.~\ref{univi}(a). Thus, it is reasonable to rely on the average values 
for the lifetime, and to derive the rapidly varying MFP as $\lambda=vt=v/V_{\rm i}$, 
where $V_{\rm i}$ is the inverse lifetime (optical potential), which enters the 
Hamiltonian through the imaginary term $-iV_{\rm i}$. For a nearly free electron, 
the wave vector $k$ acquires an imaginary part $\kappa=mV_{\rm i}/k\hbar^2$, which 
yields $\lambda=1/2\kappa$, and the time to reach the surface is then 
\begin{equation}
\tau=\frac{\hbar}{2V_{\rm i}}.
\end{equation}
     
Note that the group velocity cancels, so $\tau$ depends on energy through the 
function $V_{\rm i}(E)$. Vast empirical data suggest that the $V_{\rm i}(E)$ 
curves are similar for different materials: $V_{\rm i}$ grows with energy with 
similar rate, see Fig.~\ref{univi}(a). Because the empirically derived data may
be affected by various extrinsic factors and by the imperfection of the apparatus
it is important to theoretically corroborate this observation. Microscopically, 
optical potential is associated with the imaginary part of the self-energy. 
Figure~\ref{univi}(a) shows an {\it ab initio} $GW$ calculation for a high-energy 
conducting band of W(110). It is seen to agree well with the empirical results, 
which supports our understanding of the electron absorption as coming predominantly 
from the electron-electron interaction and corroborates the empirical estimate of 
its growth with energy. 

Thus, within this simplest approach the arrival delay of photoelectrons with 
final energy $E$ relative to those with energy $E+\epsilon$ behaves roughly 
as $\Delta\tau(E)\sim\epsilon/E(E-\epsilon)$, see Fig.~\ref{univi}(b): 
the electron with a lower energy comes later, and the delay $\Delta\tau$ 
decreases as the final state energy grows. The recent measurements by 
Neppl~\cite{NepplThesis} on W(110) of the phase shift between the spectrograms 
of valence 5$d$ and semi-core 4$f$ states qualitatively follow this 
trend [open circles in Fig.~\ref{univi}(b)], but the absolute values in the 
experiment are about three times larger. At the same time, the value of 
$5\pm 20$~as measured on Mg(0001) at $\hbar\omega=118$~eV is 4 times smaller
than expected from the simple MFP-based theory.   

\section{FUNDAMENTAL QUESTIONS\hfill ~}
Such drastic disagreement in absolute values calls for the search of  
mechanisms that, in addition to the finite lifetime, affect the transport 
of the photoelectron to the surface. A number of fundamental questions 
arise: What is the role of the finite duration of the pump pulse, i.e., 
at what point does the wave packet start its motion with the group velocity? 
Does the electron have the group velocity dictated by the band structure if 
it travels only the distance of a few atomic layers? How does the packet 
propagate when its energy falls into a forbidden gap, where the group velocity 
cannot be defined as $dE/dk$? Can the electron excited at the outermost 
atomic plane be thought of as not feeling the band structure because it 
moves away from the surface, towards vacuum?

Apart from that, an important theoretical question arises of whether the 
absorbing potential $-iV_{\rm i}$ can be used in the time-dependent
Schr\"odinger equation to adequately describe both the photocurrent 
attenuation and the implications for the transport: inelastic scattering 
is a stochastic process, so the photocurrent is an incoherent overlap of 
randomly dephased contributions, whereas the optical-potential-damped wave 
packet remains fully coherent, which may have implications both for its 
transport and for its acceleration by the laser field.

\begin{figure*}%[t]
\includegraphics[width=0.97\textwidth]{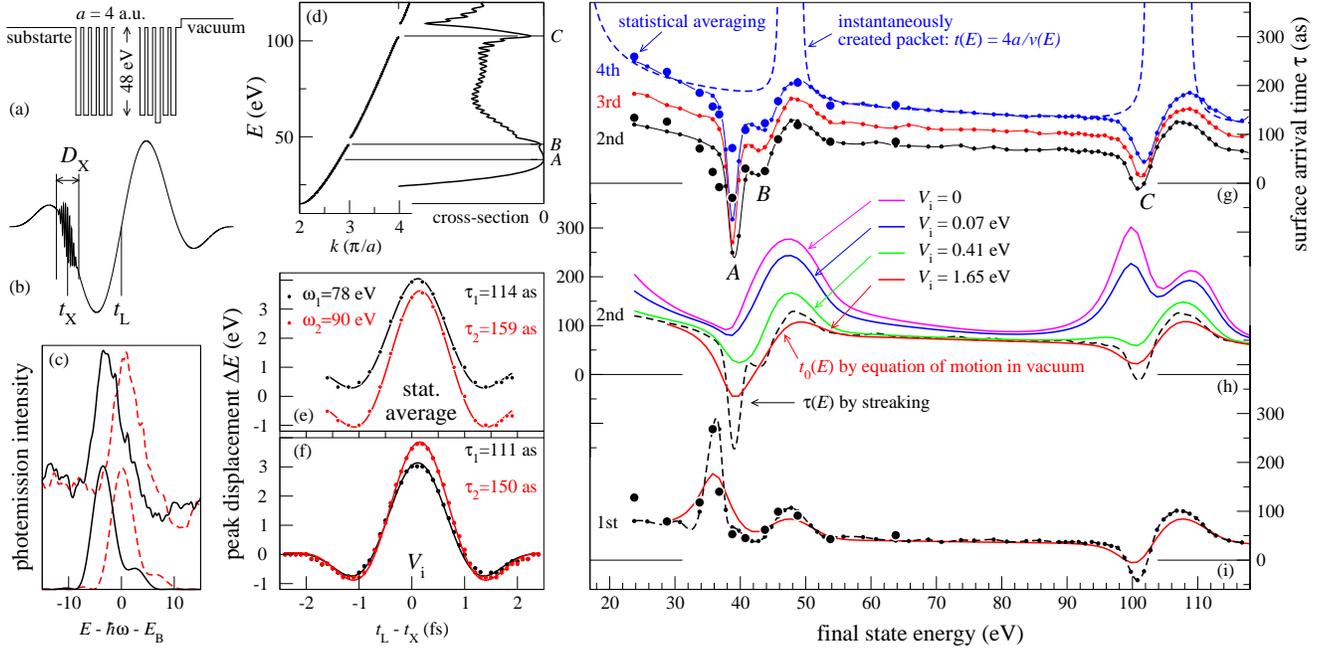}
\caption{
\label{procedure} 
(a) Crystal potential $V(z)$ with a defect at the third layer.
(b) Superposition of the XUV and the laser pulses.
(c) Streaked spectra for $\hbar\omega=78$~eV with absorption 
(lower curves) and with random collisions (upper curves) for 
$t_{\textsc l}-t_{\textsc x}=1600$~as (solid lines) and 200~as
(dashed lines). 
(d) Band structure with the periodic potential $V(z)$ and energy dependence 
of the cross-section of emission from a localized state at a defect. 
(e) and (f) Streaking curves for $\hbar\omega=78$ and 90~eV with random 
collisions (e) and with an absorption (f). (g)--(i) Surface arrival
time as function of the final state energy. Small circles connected by solid
lines in graph (g) and dashed lines in (h) and (i) are streaking results with 
absorption. Large circles in graphs (g) and (i) are streaking with random 
collisions. Solid lines in graphs (h) and (i) are from
the equation of motion in vacuum.
}
\end{figure*}
\section{THE MODEL\hfill ~}
To resolve these questions, we need a model that does not incorporate any 
assumptions about the answers and, at the same time, can be treated 
numerically exactly. For the present proof-of-concept calculation we use a 
one-dimensional model of a crystal, which treats photoexcitation, transport, 
and streaking fully quantum-mechanically without resorting to a perturbational 
treatment of any of the terms in the Hamiltonian or to a heuristic separation 
of the XUV and the laser pulses. We employ a particle-in-the-box method: the 
box comprises a thick crystal slab lying on a structureless substrate
[the piecewise constant potential $V(z)$ in Fig.~\ref{procedure}(a)] and the 
vacuum half-space. In Ref.~\cite{Krasovskii11} this model was applied to 
photoelectron streaking by a spatially uniform laser field in the absence 
of inelastic scattering.
%, and a strong influence of the band structure on the streaking spectrogram 
% was revealed. 
The present model allows for a rapid decay of the laser field into the solid 
and includes the inelastic scattering in two alternative ways: via the optical 
potential and by a straightforward ensemble averaging over random configurations. 
% The external field $E(t)$ is a superposition of an XUV and a laser pulse, see 
% Fig.~\ref{procedure}(b) with a linear polarization along the surface normal. 

\section{THE PROCEDURE\hfill ~}
We perform a series of numerical experiments, in which the system is excited 
by an XUV pulse of duration $D_{\textsc x}=1$~fs and is simultaneously acted upon 
by the laser pulse of duration $D_{\textsc l}=5$~fs, Fig.~\ref{procedure}(b). 
Both light fields are given by the dipole operator $z$, and the laser field 
is screened by multiplying it by a smooth step function $\theta(z)$. The temporal 
envelopes of both pulses are of the form $\cos^2(\pi t/D)$. The laser 
pulse is $\mathcal{E}_{\textsc l}(t)=\mathcal{E}_{\textsc l}^{\textsc m}\sin\Omega(t-t_{\textsc l})
\cos^2[\pi(t-t_{\textsc l})/D_{\textsc l}]$, with photon energy $\Omega=1.65$~eV and 
amplitude $\mathcal{E}_{\textsc l}^{\textsc m}=2\times 10^7$~V/cm. 
The TDSE is solved with the split-operator technique in matrix form in terms 
of exact eigenfunctions (discrete spectrum) of the unperturbed Hamiltonian 
$\hat H=\hat p^2/2m+V$, so the crystal potential is fully 
taken into account for both initial and final states. The spectrum of $\hat H$ 
is truncated at 200~eV above the vacuum level, which ensures the convergence 
of all the results \footnote{More details of the methodology can be found in 
Ref.~\onlinecite{KB07}, where the model was applied to a single atom.}. 

Apart from $\hat H_0$ and the two external fields the Hamiltonian includes 
inelastic scattering, which is either a static absorbing potential 
$-iV_{\rm i}[1-\theta(z)]$ or a stochastic perturbation $q(t)s_n(z)[1-\theta(z)]$, 
where $q(t)$ and $s_n(z)$ are random functions of $t$ and $z$, respectively. Here 
$q(t)$ is fixed and $s_n$ is the $n$th sample in a random ensemble. % $N_{\rm R}$ 
The ultimate spectrum is then the average over $N_{\rm R}$ random walks, 
Fig.~\ref{procedure}(c). 
The peak displacement $\Delta E$ from its laser-free position as a function of the 
time shift $\Delta t=t_{\textsc l}-t_{\textsc x}$ between the XUV and the laser pulse 
gives the key to the temporal information~\footnote{
The peak position is extracted by fitting the spectrum with a sum
of a Gaussian and a linear function to subtract the background. The background 
comes not only from the inelastically scattered photoelectrons but also from 
the random potential acting on the initial state.}: by fitting the measured 
$\Delta E(\Delta t)$ points with the momentum transfer function 
$p(\tau)=e\!\int\limits_{\tau}^{\infty}\!dt\,\mathcal{E}_{\textsc l}(t-t_{\textsc x})$, 
we infer the arrival time of the electron in vacuum $\tau$, see 
Fig.~\ref{procedure}(e) for the motion in the random potential and 
Fig.~\ref{procedure}(f) for the motion in the absorbing potential.

\section{RESULT: XUV PULSE SHIFTS PACKET \hfill ~}
For a detailed study of the photoelectron transport to the surface it is
convenient to exactly know its initial position in space. This is achieved 
by introducing a small defect at one of the layers and considering the 
photoemission from the localized state at the defect.

Figures~\ref{procedure}(e) and ~\ref{procedure}(f) show streaking curves for 
the initial state localized at the third layer for $\hbar\omega=78$ and 90~eV. 
Note that the results by the phenomenological absorbing potential (f) and by 
the microscopic random collisions (e) perfectly agree. Counterintuitively, the 
electron at the higher energy arrives 40~as later than at the lower energy:
$\tau=150$~as at 90~eV and $\tau=111$~as at 78~eV. Figure~\ref{procedure}(g) 
shows the arrival time of the electron initially at the 2nd, 3rd, and 4th layer 
for photon energies from 65 to 159~eV. The most striking are the two minima at 
$\hbar\omega=80$ and at 143~eV ($E=39$ and at 102~eV), at which the arrival 
time even shows negative values.  

\begin{figure}[t]
\includegraphics[trim = 3.0cm 1.0cm 5.2cm 2.7cm, clip=true,width=0.23\textwidth]{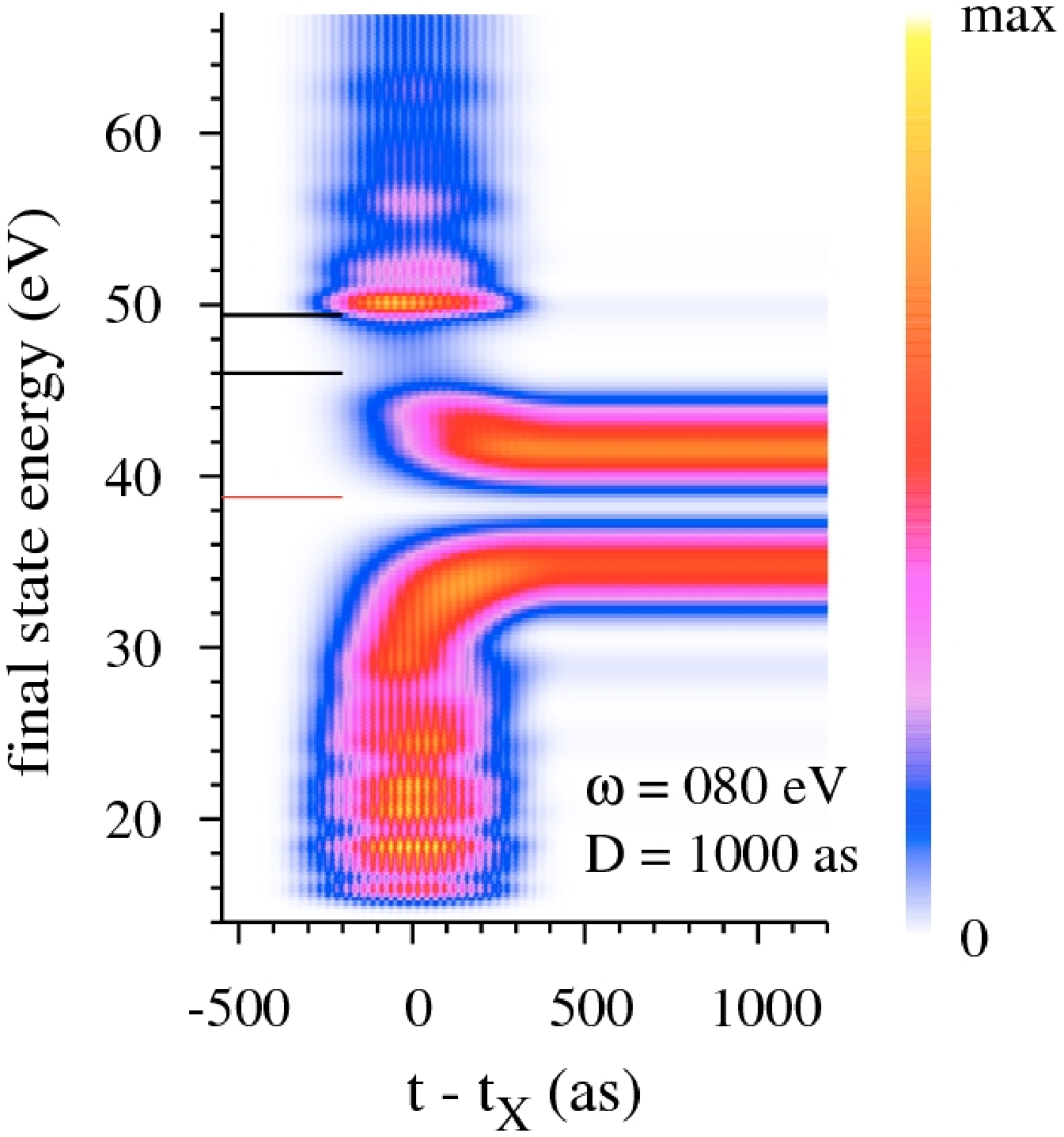}
\includegraphics[trim = 5.2cm 1.0cm 3.0cm 2.7cm, clip=true,width=0.23\textwidth]{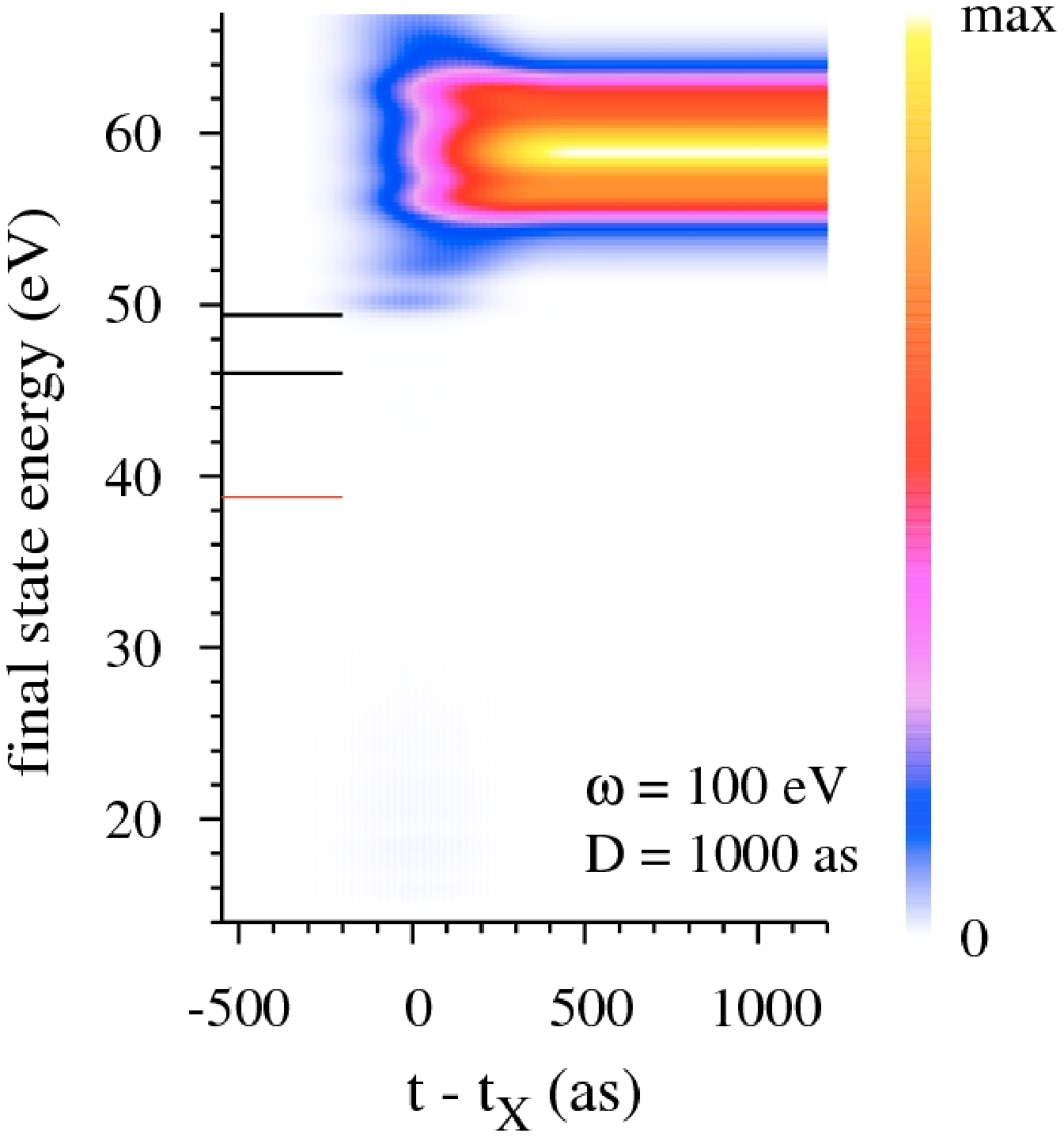}
\caption{
\label{evolution} Temporal evolution of the photoelectron spectrum from a
localized state at $\hbar\omega=80$~eV (left) and $\hbar\omega=100$~eV (right). 
The black bars at 46 and 50~eV indicate a spectral gap, and the red bar at 39~eV 
indicates the Cooper minimum.}
\end{figure}

The minima $B$ and $C$ are located at the lower edges of the band gaps, and minimum 
$A$ occurs at the energy where the ionization cross-section vanishes, see 
Fig.~\ref{procedure}(d). There the $\tau(E)$ curve strongly deviates from that 
derived from the group velocity, $t(E)=d/v(E)$ [dashed curve in Fig.~\ref{procedure}(g)]. 
To prove that the discrepancy is not due to the rather indirect streaking method to 
measure $\tau$, we employ an alternative method: we switch off the laser and follow 
the propagation of the wave packet up to a distance of 500~a.u. away from the surface. 
By measuring the center of gravity of the packet in a number of time points we determine 
its equation of motion in vacuum $z_0+\tilde{v}t$ and obtain the time point $t_0$ at 
which it has arrived at the surface. The $t_0(E)$ and $\tau(E)$ curves are seen to 
agree well, Fig.~\ref{procedure}(h), which means that the clock provided by the 
streaking technique is accurate, and that the fast delivery of the photoelectron 
to the surface is effected during the excitation process. Figure~\ref{evolution} 
shows the time evolution of the spectrum at the $\tau(E)$ minimum, $\hbar\omega=80$~eV, 
and at $\hbar\omega=100$~eV, where $\tau(E)$ well agrees with the instantaneous 
approximation $d/v(E)$. The spectral evolution maps in the two cases are qualitatively 
different: while at 100~eV the spectrum rapidly concentrates around 
$E=\hbar\omega+E_{\rm i}$, at 80~eV the evolution is much slower: at $t=200$~as 
the intensity is still spread over a 40~eV wide spectral interval. Clearly, while 
the spectral coefficients $\psi(E,t)$ of the packet $\int\!dE\,\psi(E,t)\ket{E}$ 
keep changing, the velocity of its gravity center may deviate from the weighted 
group velocity of the stationary waves $\ket{E}$, and during the XUV pulse 
the wave packet may travel a distance comparable with the mean free path.

Such unusual behavior happens every time when the central energy of
the wave packet approaches an intensity minimum, be it a vanishing
matrix element (Cooper minimum) or a spectral gap. For the emission
from the first layer, at its Cooper minimum ($\hbar\omega=77$~eV) the
XUV pulse, on the contrary, detains the packet in the solid causing a
delay of $\sim 200$~as. All the other features of the $\tau(E)$ curve
are similar to those of the emission from the deeper layers.

Figure~\ref{procedure}(h) shows that the inelastic scattering plays an important 
role in the formation of the spatial shape and transport of the packet: the 
displacement effect is more pronounced at larger $V_{\rm i}$. In this phenomenological
approach the dephasing of the wave packet is neglected, and the question arises if 
this may bring about any artifacts. This simplification turns out to be not crucial:
Figure~\ref{procedure}(g) shows the $\tau(E)$ data by statistical averaging over 144 
configurations (large circles) for emission from the 2nd and 4th layer and
Fig.~\ref{procedure}(i) for the 1st layer. The parameters of the random perturbation 
are chosen such that it gives the same MFP as $V_{\rm i}=1.7$~eV. The dephasing is
seen to cause a broadening of the $\tau(E)$ features, but otherwise the behavior of
the incoherent ensemble is in perfect accord with that of the coherent packet moving 
in the absorbing medium. That such two very different systems agree to within fine 
details suggests that this advancement phenomenon is robust, and it should be expected 
to occur in real solids.

\section{CONCLUSIONS \hfill ~}
The present calculation has revealed a peculiar dynamics of photoelectrons excited
by an ultra-short pulse. The interesting features arise from the interplay between
the elastic scattering from the crystal lattice and inelastic scattering that causes
the damping of the electron wave. Three aspects can be distinguished: first, the effect 
of inelastic scattering is to increase the average velocity of the wave packet, which 
is natural because it is the slower components that are damped stronger, as they spend 
more time in the solid. The range of the advancement effect depends on the interaction 
with the crystal lattice: at the special points, and around the gaps it may exceed 200~as, 
whereas in the free-electron-like region it reduces to 20~as, which is still comparable 
with the measured relative delays, see Fig.~\ref{univi}(b). This effect can be understood 
in terms of the group velocity of Bloch electrons. 

At certain photon energies, however, owing to a complicated non-free-electron-like
structure of the final states, the wave packet performs a complicated motion, and it
may approach the surface much faster than an instantaneously created packet. This
may also lead to an interesting effect that the electron starting from the outermost
layer is overtaken by the electron coming from the depth of the crystal. 

Close to the spectral gaps the temporal shift of the spectrogram does not necessarily
agree with the equation of motion of the center of gravity of the wave packet, but apart
from the special points discussed above the discrepancy of the two methods does not
exceed 20~as.

\vfill

\begin{acknowledgments}
This work was supported by the Spanish Ministry of Economy and Competitiveness 
MINECO (Project No. FIS2013-48286-C2-1-P).
\end{acknowledgments}

% \bibliography{ax20150607}
\end{document}